\documentclass[galaxies,article,accept,oneauthor,pdftex,10pt,a4paper]{mdpi}
\usepackage{amssymb}
\firstpage{1}
\articlenumber{85}
\issuenum{3}
\pubvolume{6}
\pubyear{2018}
\copyrightyear{2018}
\history{Received: 7 July 2018; Accepted: 3 August 2018; Published: 6 August 2018}
\updates{yes}
\Title{Surveying Planetary Nebulae Central Stars for Close Binaries: Constraining Evolution of Central Stars Based on Binary Parameters}

\Author{Todd Hillwig
\href{https://orcid.org/0000-0002-0816-1090}{\orcidicon}}

\AuthorNames{Todd Hillwig}

\address[1]{%
Department of Physics \& Astronomy, Valparaiso University, Valparaiso, IN 46383, USA; Todd.Hillwig@valpo.edu}

\corres{Correspondence: Todd.Hillwig@valpo.edu}

\abstract{The increase in discovered close binary central stars of planetary nebulae is leading to a sufficiently large sample to begin to make broader conclusions about
the effect of close binary stars on common envelope evolution and planetary nebula formation.  Herein I review some of the recent results and conclusions specifically
relating close binary central stars to nebular shaping, common envelope evolution off the red giant branch, and the total binary fraction and double degenerate
fraction of central stars.  Finally, I use parameters of known binary central stars to explore the relationship between the proto-planetary nebula and planetary
nebula stages, demonstrating that the known proto-planetary nebulae are not the precursors of planetary nebulae with close binary central~stars.}


\begin{document}
%%%%%%%%%%%%%%%%%%%%%%%%%%%%%%%%%%%%%%%%%%
%% Only for the journal Gels: Please place the Experimental Section after the Conclusions

%%%%%%%%%%%%%%%%%%%%%%%%%%%%%%%%%%%%%%%%%%
\setcounter{section}{0} %% Remove this when starting to work on the template.
%\section{How to Use this Template}
%The template details the sections that can be used in a manuscript. Note that the order and names of article sections may differ from the requirements of the journal (e.g., the positioning of the Materials and Methods section). Please check the instructions for authors page of the journal to verify the correct order and names. For any questions, please contact the editorial office of the journal or support@mdpi.com. For LaTeX related questions please contact Janine Daum at latex-support@mdpi.com.
%The order of the section titles is: Introduction, Materials and Methods, Results, Discussion, Conclusions for these journals: aerospace,algorithms,antibodies,antioxidants,atmosphere,axioms,biomedicines,carbon,crystals,designs,diagnostics,environments,fermentation,fluids,forests,fractalfract,informatics,information,inventions,jfmk,jrfm,lubricants,neonatalscreening,neuroglia,particles,pharmaceutics,polymers,processes,technologies,viruses,vision

\section{Introduction}

The study of binary stars in planetary nebulae (PNe) has the potential to provide information about the formation of PNe and the evolution of the central stars (CSs).
A number of different surveys, using different methods, have identified binary central stars of planetary nebulae (CSPNe).  Photometric surveys looking for
variability due to a companion tend to find close binary systems with orbital periods of about a week or less (e.g., \cite{bon90,mis09,dem15}).  Such photometric
surveys are sensitive to both cool companions and double degenerate systems \cite{hil11,san11}.  Infrared surveys are designed to be sensitive to cool companions at all
orbital periods \cite{dou15}, and radial velocity surveys are sensitive to stellar mass companions in orbital periods of up to several years \cite{dem04,jon17}.

Once binary CSPNe are identified, follow-up work can determine system parameters.  A number of such systems have recently been studied, with various sets of orbital
and stellar parameters published (see the updated list of known close binary CSPNe maintained by David Jones at \url{http://drdjones.net/?q=bCSPN}).
Because it is more difficult to determine parameters for binaries with long periods, the majority of
binary CSPNe with known physical parameters are close binaries.  Of the close binary CSPNe, the large majority have been discovered using photometry.

In addition to providing discovery data for binaries, the light curves of those binaries can also tell us a great deal about the nature of the binary system.  For example,
close binary CSPNe with a main sequence companion have light curves dominated by an irradiation effect in which the inner hemisphere of the cool star is irradiated
and heated by the hot CS.  This behavior results in a nearly sinusoidal light curve with one maximum and one minimum per orbit.  However, if the companion to the
hot CS is a compact object (e.g., a white dwarf, WD), then any detected variability will likely be through eclipses or, more likely, ellipsoidal variability that
is due to one of the stars nearly filling its Roche lobe and thus being elongated by the mutual gravity of the two stars.  Typically, it will be the CS that nearly fills
its Roche lobe, since these objects can still be large, not having contracted yet to the WD cooling track.  As the CS contracts, the ellipsoidal variability decreases
in amplitude, eventually becoming unobservable. A light curve dominated by ellipsoidal variability has two maxima
and two minima per orbit due to the different projected surface areas (brighter when seen edge-on, fainter when the two stars are aligned along the line of sight).
To date, all well-studied close binary CSPNe with light curves dominated by ellipsoidal variability have been shown to be double
degenerate (DD) systems.  Technically, the CS is typically not yet fully degenerate, but we use this terminology here to provide a better comparison with true DD systems,
which these will become.  They are also related to core degenerate (CD) systems \cite{kas11}, though in this case the common envelope (CE) has detached and left a close
binary system.

With the growing number of discovered binary CSPNe, we are beginning to reach a point at which statistically relevant statements can be made about
how binary companions may influence both the ejection of the nebula and the evolution of the central star.  Below I review some conclusions
that have already been discussed in the literature, along with several connections that are currently being explored but still need confirmation.
I then discuss what our current knowledge of close binary CSPNe
can tell us about the evolution of PNe with close binary nuclei, especially in the context of proto-planetary nebulae (PPNe).

%%%%%%%%%%%%%%%%%%%%%%%%%%%%%%%%%%%%%%%%%%
\section{Relationships between Close Binary CSs and Their PNe}
\unskip
\subsection{The Connection between Central Binaries and PN Shaping}

One of the earliest predictions of close binary evolution via common envelope evolution was by Paczynski \cite{pac76}.  Shortly afterward, discussions began
about how important CE evolution might be to the ejection of stellar envelopes and the production of PNe \cite{bon78}.  These discussions led to exploring the
relationship between binary interactions and the non-spherical structures of most PNe (e.g., \cite{mor81,bon90}).  A~more recently described option is
grazing envelope evolution (GEE), in which the binary orbit shrinks as mass is lost from a giant star via Roche lobe overflow.  However,
jet launching from the accreting star strips away enough of the overflowing mass to prevent the CE \cite{sok15}.

Even though it seems intuitive that CE or GEE evolution could, and perhaps would, affect the shape of the ejected envelope, the details of how this happens are
not fully understood.  Our knowledge of CE evolution has seen a dramatic increase over the past five years or so, but as a computational problem we still
do not understand the full process from spiral-in, through ejection, and on to a visible PN \cite{iva18}.  GEE is a more recent theory and does not have the
volume of work performed at this time, and it is not clear that it can result in short-period systems with separations of a few $R_\odot$ or less \cite{abu18}.
For this reason, the remainder of this work will focus on CE evolution, but the reader should take into account that GEE could also play a role.

However, enough systems with known binary inclination and nebular inclination now exist to demonstrate a direct relationship between PN shapes and close
central binaries \cite{hil16b}.  The clear observational connection between binary CSs and their PNe combined with computational work (e.g.,~\cite{nor17})
allows us to conclude that the close binary CSs either strongly affect, or completely determine, the shape of the PN.

Binary CSs also appear to be connected to PNe with high abundance discrepancy factors (ADFs).
ADFs are the result of very different abundance values calculated from collisional versus recombination line strengths.  Recent work
\cite{cor15,jon16} shows a strong correlation between high ADF and binary CSs in PNe.  In addition, there is evidence in at least one case
that the ADF is caused by two clearly distinct spatial structures, such that the abundance differences result from different physical
parameters in the two spatially distinct regions \cite{jon16,gar16}.  These regions may well be another example of PN shaping resulting from binary interactions.

\subsection{The CE Phase and Evolution of the CS}

CE evolution in this context begins when one of the two stars in a binary system moves off the main sequence and expands to the point of coming into contact
with its Roche lobe with a deep convective envelope.
Actual spiral-in occurs later in the process, but the CE phase is triggered by the Roche lobe of the more massive star and its expansion off the main sequence.
For binaries that start off relatively close, the CE interaction may take place before the expanding star reaches the AGB, resulting in post-RGB objects.
If is not known if CE evolution on the RGB can result in a visible PN.  Typically,
the~slower evolution of post-RGB stars would mean that the envelope disperses before the core becomes hot enough to ionize the surrounding gas.  However, Hall et al.
\cite{hal13} discuss some circumstances under which a post-RGB core may be able to ionize a visible PN.

Until recently, no CSPNe had been identified that were clearly consistent with post-RGB evolution.  Hillwig et al. \cite{hil17} provide a list of five CSs that are
potential post-RGB objects, four of which are in known close binary systems (it is possible that the fifth, HaWe~13 \cite{fre16}, is a close binary, but no direct evidence
currently exists).  Of the five, the most secure candidate for a post-RGB CS is that of ESO~330-9.  Modeling of the CS spectrum and binary modeling are
both consistent with a low-mass ($M_{CS} = 0.38-0.45$ $M_\odot$) post-RGB object \cite{hil17}.

While it is not clear that any of these PNe harbor post-RGB CSs, confirming the nature of these stars, and other potential post-RGB stars
\cite{hil17} will have important implications for our understanding of CE evolution and PN production.

\subsection{Companions to Binary CSPNe and the Close Binary Fraction}

As noted above, the light curves of close binary CSPNe can give us information about the most-likely companion to the CS.  A light curve dominated by ellipsoidal
variability in all well-studied cases so far have been shown to be double degenerate (DD) systems, where the companion to the hot CS is a compact object---most likely a WD.  Likewise, light curves dominated by an irradiation effect have been shown to have cool main sequence stars as companions.  This makes
physical sense due to the nature of the system.  Because the CS is very hot,
any main sequence companion will exhibit a strong irradiation effect.  If the cool companion is large enough to nearly fill its Roche lobe, and thus exhibit
ellipsoidal variability, the resulting irradiation effect would be far stronger than the ellipsoidal variability.  Thus, for a cool companion we expect the light
curve to be dominated by irradiation.  The~combination of temperatures and luminosities mean the irradiation effect should be visible for the entire PN lifetime.

Of the currently known close binary CSPNe in the literature (for a current list, see the online table maintained by David Jones at \url{http://drdjones.net/?q=bCSPN}),
48 show photometric variability and are firmly identified as the CS of a PN.  Of those, 13 have light curves dominated by ellipsoidal variability.  If we take all of those
to be DD systems, then we find that roughly one-quarter of all detected close binary CSPNe have a WD companion.

The caveat is that these are the detected systems.  Ellipsoidal variability will only be observable while one or both stars fill a significant fraction of their Roche lobe.
In each of the well-studied cases it appears that the deformed star is the CS.  So as the CS contracts toward the WD track, it~will detach farther
from its Roche lobe, meaning the ellipsoidal variability will decrease relatively rapidly.  It~is~unclear how long the ellipsoidal variability  for a typical binary CSPN will be observable.  The author and collaborators are working on determining a realistic result for the duration of the observable ellipsoidal variability in these systems.  A
back-of-the-envelope approach results in a fractional observable lifetime (relative to the length of time the PN is visible) of about one-third to one-quarter.
This is based on the observable lifetime of PNe relative to the length of time the CS will be large enough to nearly fill its Roche lobe and display ellipsoidal
variability.  So, if ellipsoidal variability is only visible for about one-third to one-quarter of the lifetime of a PN, then we can assume that we only observe about
one-third to one quarter of all DD systems to have photometric variability.

With these pieces of information, we would expect there to be roughly as many DD close binary CSPNe as there are those with a main sequence companion.
If we use close binary fractions from the literature for photometric studies \cite{bon00,mis09} of about 10--20\%, or the value from the author's recent survey (paper in
preparation) of $\sim$12\%, and assume that we have only observed roughly one-quarter to one-third of the close DD systems, we arrive at a corrected
\emph{close} binary fraction of $\sim$20--30\%.  With~an~improved observable lifetime of ellipsoidal variability, and with a more precise value of the close binary
fraction, the uncertainty will shrink, with a goal of identifying the close binary percentage to within a 5\% range in the near future.

%%%%%%%%%%%%%%%%%%%%%%%%%%%%%%%%%%%%%%%%%%
\section{The Relationship between PPNe and PNe}

The discovery of objects in the intermediate evolutionary step between the end of the AGB and the PN phase began the study of PPNe \cite{par86,hri88,van89}.
Many PPNe
show very strongly aspherical structures~\cite{hri99,uet00,sah03}, so discussions of PN shaping naturally involved PPNe, and it seemed clear that the mechanism that shapes
PNe must act very early, such that bipolar structures are apparent even in the early PPN stages.

Various photometric, spectroscopic, and radial velocity studies of PPNe to search for close binaries that may be responsible for shaping have all ended with the absence
of any evidence for close binary cores \cite{hri07,hri17}, or in fact for any binary systems with orbital periods less than years.

The high luminosities and low temperatures of the CSs of PPNe mean that they must be relatively large---much larger than the orbital separation of any of the known
close binary CSPNe.
The distances to PPNe are not well known, though Gaia data will hopefully provide answers for some PPNe.  Luminosities
are difficult to determine without well-known distances,  though typical luminosities are expected to be in the several thousands of $L_\odot$.  With typical temperatures determined from spectra
of 5000--10,000~K, we can estimate radii of the PPN CSs to be tens of $R_\odot$ or more.  With typical orbital separations for the known close binary CSPNe of
a few $R_\odot$ or less,
for these PPNe to be the progenitors of the close binaries, they would have to still be in the CE phase.  However, current studies show that the CE
phase lasts on the order of weeks, or at most years, depending on when we begin to consider the system as entering the CE \cite{gal17}.
These PPN CSs are clearly not evolving on that timescale.

\begin{figure}[H]
\centering
\includegraphics[width=8 cm, angle=270]{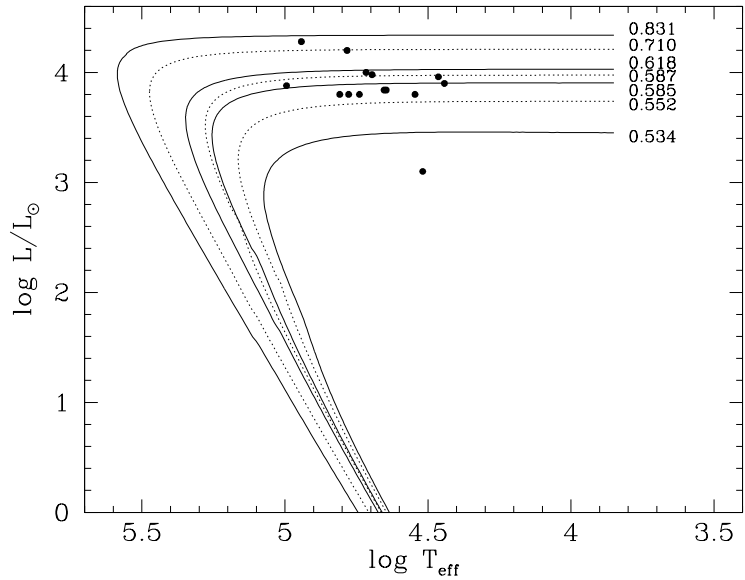}
\caption{Evolutionary tracks from Table 3 in Miller Bertolami \cite{mil16} for central stars of planetary nebulae (CSPNe) with $Z=0.01$ and for final central star (CS) masses as labeled on the upper-right next to
each track.  Line styles alternate between sold and dashed to make them easier to follow on the plot.
Also shown are the locations of the hot CS of a number of close binary CSPNe at their maximum possible radius
(see text for description).\label{fig1}}
\end{figure}

Indeed, Figure \ref{fig1} shows evolutionary tracks from Miller Bertolami \cite{mil16} along with the \emph{beginning} positions of the CSs of a number of well-studied
close binary CSPNe.  These beginning positions, the~locations of the points in Figure \ref{fig1}, are defined as the location along the CS evolutionary track where the CS
would just fill its Roche lobe, the idea being that if the CS were any larger, the system would still be in the CE phase (or at least would be undergoing significant
mass transfer).

The placement of each data point in Figure \ref{fig1} was determined by choosing a mass value for the two stars in each binary from the literature (often from
a range of values, or on occasion choosing between two or more different values from different studies).  As such, the mass values in Table \ref{tab1} \emph{should not} be
reproduced as a list of known masses---they are representative values.

The luminosity of each CS was then assigned based on its mass, using the models
of Miller Bertolami \cite{mil16}.  Each CS was assigned a luminosity approximately equal to that for a CS of that mass, at a post-AGB age of zero years.

The temperature of the CS was determined by first calculating the volume radius of the Roche lobe of the CS using the method of Eggleton \cite{egg83} where
the separation, $a$, can be found from $P_{orb}$ and the masses using Kepler's third law.  Once the radius is calculated, the temperature is approximated
from the luminosity and radius using the Stefan--Boltzmann equation.

Table \ref{tab1} shows the physical parameters from the literature and the calculated values for each~CSPN.

\vspace{-12pt}

\begin{table}[H]
\caption{Literature data for binary CSPNe along with calculated maximum radii and effective temperatures for the CSs at a luminosity appropriate for the CS mass.
\label{tab1}}
\centering
%% \tablesize{} %% You can specify the fontsize here, e.g.,  \tablesize{\footnotesize}. If commented out \small will be used.
\begin{tabular}{cccccccc}
\toprule
\textbf{PN Name}	& \boldmath{$P_{orb}$} (\textbf{days})   & \boldmath{$M_{CS}$ ($M_\odot$)}	& \boldmath{$M_2$ ($M_\odot$)} & \boldmath{$R_{max}$ ($R_\odot$)}	&
\boldmath{$\log L/L_{\odot}$}	& \boldmath{$\log T_{eff}$}	& \textbf{Ref. }\\
\midrule
V458 Vul    	&0.06812255   		&0.58    	&1          	&0.27                & 3.88       & 4.99	 & \cite{rod10}  \\
TS 01       		&0.163508     		&0.80     	&0.54       	&0.57                & 4.28       & 4.94	 & \cite{tov10}  \\
NGC 6337    	&0.1736133    		&0.56    	&0.35       	&0.53                & 3.80       & 4.81	 & \cite{hil16b}  \\
Abell 41    	&0.226453066  	&0.56    	&0.56       	&0.61                & 3.80       & 4.78	 & \cite{bru01,shi08b}  \\
HaTr 7     		&0.3221246     		&0.56    	&0.14       	&0.88                & 3.80       & 4.74	 & \cite{hil17}  \\
DS 1        		&0.35711296    		&0.70     	&0.3        	&0.96                & 4.20       & 4.78	 & \cite{dri85,hil96}  \\
Abell 63    	&0.46506921   		&0.63   	&0.29       	&1.10                & 4.00       & 4.72	 & \cite{afs08}  \\
Abell 46    	&0.47172909   		&0.51   	&0.14       	&1.09                & 3.10       & 4.51	 & \cite{afs08}  \\
Lo 16       		&0.48626      		&0.6     	&0.4        	&1.08                & 3.98       & 4.69	 & in prep
\\
NGC 6026    	&0.528086     		&0.57    	&0.57       	&1.09                & 3.84       & 4.65	 & \cite{hil10}  \\
HFG 1       	&0.5816475    		&0.57    	&1.09       	&1.13                & 3.84       & 4.65	 & \cite{shi04}  \\
Abell 65    	&1.0037577    		&0.56    	&0.22       	&1.79                &  3.80      & 4.54	 & \cite{hil16a}  \\
LTNF 1     		&2.2911667		& 0.59	& 0.25	& 3.14		& 3.96	& 4.46	 & \cite{shi08c}  \\
Sp 1       		&2.90611        		&0.58    	&0.48	& 3.45               & 3.90    	&  4.44	 & \cite{hil16b}  \\
\bottomrule
\end{tabular}
\end{table}

Figure \ref{fig1} demonstrates that these ``close'' binaries are in fact close enough that the core of the evolving star could not have started its post-AGB evolution
at the zero-age point (the top right of the plot, near the mass values) on the evolutionary tracks without over-filling its Roche lobe.
The~CSs of PPNe are observed to be single objects and stable over at least several decades through direct observation (though many show pulsations).
They do not show evidence of binarity, significant mass transfer, or large-scale changes, and the circumstellar envelope has detached from the CS \cite{hri17}.
If these objects entered a CE in the past, they are now at a stage after the CE has been ejected.
This has several implications.  First, the particular PNe with close binary CSs shown in Figure \ref{fig1} would \emph{not} have gone through the PPN phase as we know it.
Observed PPNe have CSs that are too cool and luminous (thus too large) to exist in a close binary system unless they are undergoing mass transfer or
are in the CE phase.  We also now know that the CE phase evolves rapidly to ejection of the envelope (or merger of the cores)---faster than the directly observed
lifetimes of these PPNe.  Indeed, for realistic stellar masses, none of the known close binary systems (with $P_{orb}\lesssim7$ days) could begin their
post-CE evolution at a stage (the zero-age post-AGB point) that would result in an object similar to observed PPN CSs.

One caveat to this is the possibility of a self-regulated CE phase, which may result in a longer-lived CE \cite{iva13}.  This phase is not well understood in terms of the rate of
occurrence or observable properties, but it seems unlikely that it would result in the well-behaved stellar-like properties of the observed PPN CSs.  This is especially
true given the consistent evolution of the studied PPNe, such that they would either all be undergoing self-regulated CE evolution, or it becomes even less likely
that any of them are.

So it seems clear that the known PPNe are not progenitors of PNe with close binary CSs, despite exhibiting strong nebular shaping.  Note, however, that there could
still be a wider binary companion that could engulf the current CS in a CE phase later in its evolution, producing a DD system.  However, none of the known PPNe
will proceed directly to the PN phase with a close binary CS.  At present there are several alternative possibilities for the observed morphology of PPNe:
the shaping is due to a companion (likely of low mass) that has been disrupted or has merged with the CS of the PPN, the~shaping is
caused by a wider companion, or the shaping is unrelated to a companion.

The second implication of Figure \ref{fig1} is that the binary companion clearly affects the evolution of the CS.  If the CSs of the binary systems in Table \ref{tab1} had evolved as
single stars, we would expect them to begin their evolution as a CSPN roughly at the zero-age point on the evolutionary tracks in Figure \ref{fig1}.  However, due to their
binary companions, they either need to start their evolution at a more evolved stage (higher temperature) on those evolutionary tracks, or they would over-fill their
Roche lobe and transfer mass onto their companion.  Either results in a change relative to single star evolution, or even CE evolution that would result in a
wider final separation.
Determining the result of the change in evolution is beyond the scope of the present
work, but understanding that process may provide a key element in the energetics of CE evolution.  For example, the core evolution may be accelerated as the
companion spirals in, the final mass of the CS may differ from that of an equivalent single star, the~core may undergo significant non-thermal equilibrium
evolution, or some combination of these may occur.  However, it seems that these interactions are likely to affect the overall energy budget of the CE evolution
(though whether it is a significant effect is also a question).

%% If the documentclass option "submit" is chosen, please insert a blank line before and after any math environment (equation and eqnarray environments). This ensures correct linenumbering. The blank line should be removed when the documentclass option is changed to "accept" because the text following an equation should not be a new paragraph.

%%%%%%%%%%%%%%%%%%%%%%%%%%%%%%%%%%%%%%%%%%
\section{Discussion}

I have summarized some of the recent conclusions about the relationship between close binary CSs and their PNe.  These conclusions are possible due to the
increase in the number of discovered close binary CSPNe and by the increase in the number of systems with full binary modeling.  I also demonstrated that the
PPN-to-PN connection is not as clear as it may seem.  The observed sample of PPNe are \emph{not} progenitors of PNe with close binary stars.  In addition, close
binary CSPNe seem to alter the post-AGB evolution of the core of the evolving star, with possible implications for CE evolution in addition to PN shaping and close
binary evolution.

%%%%%%%%%%%%%%%%%%%%%%%%%%%%%%%%%%%%%%%%%%
\vspace{6pt}
%%%%%%%%%%%%%%%%%%%%%%%%%%%%%%%%%%%%%%%%%%
\funding{This material is based upon work supported by the National Science Foundation under Grant No. AST-1109683. Any opinions, findings, and conclusions or recommendations expressed in this material are those of the author(s) and do not necessarily reflect the views of the National Science Foundation.}

%%%%%%%%%%%%%%%%%%%%%%%%%%%%%%%%%%%%%%%%%%
%\acknowledgments{}

%%%%%%%%%%%%%%%%%%%%%%%%%%%%%%%%%%%%%%%%%%
\conflictsofinterest{The author declares no conflict of interest.}

%=====================================
% References, variant A: internal bibliography
%=====================================
\reftitle{References}

%%%%%%%%%%%%%%%%%%%%%%%%%%%%%%%%%%%%%%%%%%

\begin{thebibliography}{999}
% Reference 1
\bibitem[Bond and Livio(1990)]{bon90}
Bond, H.E.; Livio, M. Morphologies of Planetary Nebulae Ejected by Close-Binary Nuclei. \textit{
Astrophys. J.} \textbf{1990}, \textit{355}, 568--576. [\href{http://dx.doi.org/10.1086/168789}{CrossRef}]

\bibitem[Miszalski, et al.(2009)]{mis09}
Miszalski, B.; Acker, A.; Moffat, A.F.J.; Parker, Q.A.; Udalski, A. Binary planetary nebulae
nuclei towards the Galactic bulge. I. Sample discovery, period distribution, and
binary fraction. \textit{Astron. Astrophys.} \textbf{2009}, \textit{496}, 813--825. [\href{http://dx.doi.org/10.1051/0004-6361/200811380}{CrossRef}]

\bibitem[De Marco, et al.(2015)]{dem15}
De Marco, O.; Long, J.; Jacoby, G.H.; Hillwig, T.; Kronberger, M.; Howell, S.B.; Reindl, N.; Margheim, S. Identifying close binary
central stars of PN with Kepler. \textit{Mon. Not. R. Astron.
Soc.} \textbf{2015}, \textit{448}, 3587--3602. [\href{http://dx.doi.org/10.1093/mnras/stv249}{CrossRef}]

\bibitem[Hillwig(2011)]{hil11}
Hillwig, T.C. The physical characteristics of binary central stars of planetary nebulae. In Proceedings of the Asymmetric Planetary Nebulae 5 Conference, Bowness-on-Windermere, UK, 20--25 June 2010; p. 275.

\bibitem[Santander-Garcia, et al.(2011)]{san11}
Santander-Garcia, M.; Rodr{\'\i}guez-Gil, P.; Jones, D.; Corradi, R.L.M.; Miszalski, B.; Pyrzas, S.; Rubio-D{\'i}ez,~M.M. The binary central stars of PNe with the shortest orbital period. In Proceedings of the Asymmetric
Planetary Nebulae 5 Conference, Bowness-on-Windermere, UK, 20--25 June 2010; p. 259.

\bibitem[Douchin, et al.(2015)]{dou15}
Douchin, D.; De Marco, O.; Frew, D.J.; Jacoby, G.H.; Jasniewicz, G.; Fitzgerald, M.; Passy, J.C.; Harmer,~D.; Hillwig, T.; Moe, M. The binary fraction of
planetary nebula central stars-II. A larger sample and improved technique for
the infrared excess search. \textit{Mon. Not. R. Astron. Soc.}
\textbf{2015}, \textit{448}, 3132--3155. [\href{http://dx.doi.org/10.1093/mnras/stu2700}{CrossRef}]

\bibitem[De Marco, et al.(2004)]{dem04}
De Marco, O.; Bond, H.E.; Harmer, D.; Fleming, A.J. Indications of a Large Fraction of Spectroscopic
Binaries among Nuclei of Planetary Nebulae. \textit{Astrophys. J.} \textbf{2004}, \textit{602},
L93. [\href{http://dx.doi.org/10.1086/382156}{CrossRef}]

\bibitem[Jones, et al.(2017)]{jon17}
Jones, D.; Van Winckel, H.; Aller, A.; Exter, K.; De Marco, O. The long-period binary central stars
of the planetary nebulae NGC 1514 and LoTr 5. \textit{Astron. Astrophys.} \textbf{2017},
\textit{600}, L9. [\href{http://dx.doi.org/10.1051/0004-6361/201730700}{CrossRef}]

\bibitem[Kashi and Soker(2011)]{kas11}
Kashi, A.; Soker, N. A circumbinary disc in the final stages of common envelope and the core-degenerate
scenario for Type Ia supernovae. \textit{Mon. Not. R. Astron. Soc.} \textbf{2011}, \textit{417}, 1466--1479. [\href{http://dx.doi.org/10.1111/j.1365-2966.2011.19361.x}{CrossRef}]

\bibitem[Paczynski(1976)]{pac76}
Paczynski, B. Common Envelope Binaries. \textit{Struct. Evol. Close Bin. Syst.} \textbf{1976}, \textit{73}, 75.

\bibitem[Bond, et al.(1978)]{bon78}
Bond, H.E.; Liller, W.; Mannery, E.J. UU Sagittae: Eclipsing nucleus of the planetary nebula Abell 63.
\textit{Astrophys. J.} \textbf{1978}, \textit{223}, 252. [\href{http://dx.doi.org/10.1086/156257}{CrossRef}]

\bibitem[Morris(1981)]{mor81} Morris, M. Models for the structure
and origin of bipolar nebulae. \textit{Astrophys. J.} \textbf{1981}, \textit{249}, 572--585. [\href{http://dx.doi.org/10.1086/159317}{CrossRef}]

\bibitem[Soker(2015)]{sok15}
Soker, N. Close Stellar Binary Systems by Grazing Envelope Evolution. \textit{Astrophys. J.} \textbf{2015}, \textit{800}, 114. [\href{http://dx.doi.org/10.1088/0004-637X/800/2/114}{CrossRef}]

\bibitem[Ivanova and Nandez(2018)]{iva18}
Ivanova, N.; Nandez, J. Planetary Nebulae Embryo after a Common Envelope Event. \textit{Galaxies}
\textbf{2018}, \textit{6}, 75. [\href{http://dx.doi.org/10.3390/galaxies6030075}{CrossRef}]

\bibitem[Abu-Backer, et al.(2018)]{abu18}
Abu-Backer, A.; Gilkis, A.; Soker, N. Orbital Radius during the Grazing Envelope Evolution. \textit{Astrophys. J.} \textbf{2018}, \textit{861}, 136. [\href{http://dx.doi.org/10.3847/1538-4357/aacb77}{CrossRef}]

\bibitem[Hillwig, et al.(2016)]{hil16b}
Hillwig, T.C.; Jones, D.; De Marco, O.; Bond, H.E.; Margheim, S. Frew, D. Observational
Confirmation of a Link between Common Envelope Binary Interaction and Planetary
Nebula Shaping.  \textit{Astrophys. J.} \textbf{2016}, \textit{832}, 125,
doi:10.3847/0004-637X/832/2/125. [\href{http://dx.doi.org/10.3847/0004-637X/832/2/125}{CrossRef}]

\bibitem[Nordhaus(2017)]{nor17}
Nordhaus, J. \textit{Planetary Nebulae: Multi-Wavelength Probes of Stellar and Galactic Evolution};  IAU Symposium;  Cambridge University Press: Cambridge, UK, 2017; Volume 323, p. 207, doi:10.1017/S1743921317002216.

\bibitem[Corradi, et al.(2015)]{cor15}
Corradi, R.L.M.; Garc{\'\i}a-Rojas, J.; Jones, D.; Rodr{\'\i}guez-Gil, P. Binarity and the
Abundance Discrepancy Problem in Planetary Nebulae. \textit{Astrophys. J.} \textbf{2015}, \textit{803}, 99. [\href{http://dx.doi.org/10.1088/0004-637X/803/2/99}{CrossRef}]

\bibitem[Jones, et al.(2016)]{jon16}
Jones, D.; Wesson, R.; Garc{\'\i}a-Rojas, J.; Corradi, R.L.M.; Boffin, H.M.J. NGC 6778:
Strengthening the link between extreme abundance discrepancy factors and central
star binarity in planetary nebulae. \textit{Mon. Not. R. Astron. Soc.} \textbf{2016}, \textit{455}, 3263--3272. [\href{http://dx.doi.org/10.1093/mnras/stv2519}{CrossRef}]

\bibitem[Garc{\'\i}a-Rojas, et al.(2016)]{gar16}
Garc{\'\i}a-Rojas, J.; Corradi, R.L.M.; Monteiro, H.; Jones, D.;
Rodr{\'\i}guez-Gil, P. Cabrera-Lavers, A. Imaging the Elusive H-poor Gas in
the High adf Planetary Nebula NGC 6778. \textit{Astrophys. J.} \textbf{2016}, \textit{824},
L27. [\href{http://dx.doi.org/10.3847/2041-8205/824/2/L27}{CrossRef}]

\bibitem[Hall, et al.(2013)]{hal13}
Hall, P.D.; Tout, C.A.; Izzard, R.G.; Keller, D. Planetary nebulae after common-envelope phases
initiated by low-mass red giants. \textit{Mon. Not. R. Astron.
Soc.} \textbf{2013}, \textit{435}, 2048--2059. [\href{http://dx.doi.org/10.1093/mnras/stt1422}{CrossRef}]

\bibitem[Hillwig, et al.(2017)]{hil17}
Hillwig, T.C.; Frew, D.J.; Reindl, N.; Rotter, H.; Webb, A. Margheim, S. Binary Central
Stars of Planetary Nebulae Discovered through Photometric Variability. V. The
Central Stars of HaTr 7 and ESO 330-9. \textit{Astrophys.~J.} \textbf{2017}, \textit{153}, 24,
doi:10.3847/1538-3881/153/1/24. [\href{http://dx.doi.org/10.3847/1538-3881/153/1/24}{CrossRef}]

\bibitem[Frew, et al.(2016)]{fre16}
Frew, D.J.; Parker, Q.A.; Boji{\v{c}}i{\'c}, I.S. The H{\ensuremath{\alpha}} surface brightness-radius relation: A robust statistical distance indicator for planetary nebulae.
\textit{Mon. Not. R. Astron. Soc.} \textbf{2016}, \textit{455}, 1459--1488,
doi:10.1093/mnras/stv1516. [\href{http://dx.doi.org/10.1093/mnras/stv1516}{CrossRef}]

\bibitem[Bond(2000)]{bon00}
Bond, H.E. Binarity of Central Stars of Planetary Nebulae. \textit{arXiv} \textbf{1999}, arXiv:astro-ph/9909516.

\bibitem[Parthasarathy and Pottasch(1986)]{par86}
Parthasarathy, M.; Pottasch, S.R. The far-infrared (IRAS) excess in HD
161796 and related stars. \textit{Astron.~Astrophys.} \textbf{1986}, \textit{154}, L16.

\bibitem[Hrivnak, et al.(1989)]{hri88}
Hrivnak, B.J.; Kwok, S.; Volk, K.M. The High-Latitude F Supergiant IRAS 18095+2704: A Proto-Planetary Nebula.  \textit{Astrophys. J.} \textbf{1988}, \textit{331}, 832--837. [\href{http://dx.doi.org/10.1086/166603}{CrossRef}]

\bibitem[van der Veen, et al.(1989)]{van89}
Van der Veen, W.E.C.J.; Habing, H.J.; Geballe, T.R. Objects in transition from the
AGB to the planetary nebula stage: New visual and infrared observations.
\textit{Astron. Astrophys.} \textbf{1989}, \textit{226}, 108.

\bibitem[Hrivnak, et al.(1999)]{hri99}
Hrivnak, B.J.; Langill, P.P.; Su, K.Y.L.; Kwok, S. Subarcsecond Optical Imaging of Proto-Planetary Nebulae. \textit{Astrophys. J.} \textbf{1999}, \textit{513}, 421. [\href{http://dx.doi.org/10.1086/306821}{CrossRef}]

\bibitem[Ueta, et al.(2000)]{uet00}
Ueta, T.; Meixner, M.; Bobrowsky, M. A Hubble Space Telescope Snapshot Survey of Proto-Planetary Nebula
Candidates: Two Types of Axisymmetric Reflection Nebulosities. \textit{Astrophys. J.} \textbf{2000}, \textit{528}, 861. [\href{http://dx.doi.org/10.1086/308208}{CrossRef}]

\bibitem[Sahai(2003)]{sah03}
Sahai, R. Multi-polar Structures in Young Planetary and Protoplanetary Nebulae. In \textit{Symposium-International Astronomical Union}; Cambridge University Press: Cambridge, UK, 2003; Volume 209, p. 471.

\bibitem[Hrivnak(2007)]{hri07}
Hrivnak, B.J. A Search for Binaries in Proto-Planetary Nebulae. In Proceedings of the Asymmetrical Planetary Nebulae IV, La Palma, Spain, 18--22 June 2017.

\bibitem[Hrivnak, et al.(2017)]{hri17}
Hrivnak, B.J.; Van de Steene, G.; Van Winckel, H.; Sperauskas, J.; Bohlender, D. Lu, W. Where
are the Binaries? Results of a Long-term Search for Radial Velocity Binaries in
Proto-planetary Nebulae. \textit{Astrophys. J.} \textbf{2017}, \textit{846}, 96,
doi:10.3847/1538-4357/aa84ae. [\href{http://dx.doi.org/10.3847/1538-4357/aa84ae}{CrossRef}]

\bibitem[Galaviz, et al.(2017)]{gal17}
Galaviz, P.; De Marco, O.; Passy, J.-C.; Staff, J.E.; Iaconi, R. Common Envelope Light Curves. I.
Grid-code Module Calibration. \textit{Astrophys. J. Suppl. Ser.} \textbf{2017},
\textit{229}, 36. [\href{http://dx.doi.org/10.3847/1538-4365/aa64e1}{CrossRef}]

\bibitem[Miller Bertolami(2016)]{mil16}
Miller Bertolami, M.M. New models for the evolution of post-asymptotic giant branch stars and central
stars of planetary nebulae. \textit{Astron. Astrophys.} \textbf{2016}, \textit{588}, A25,
doi:10.1051/0004-6361/201526577. [\href{http://dx.doi.org/10.1051/0004-6361/201526577}{CrossRef}]

\bibitem[Eggleton(1983)]{egg83}
Eggleton, P.P. Aproximations to the radii of Roche lobes.  \textit{Astrophys. J.} \textbf{1983}, \textit{268}, 368,
doi:10.1086/160960. [\href{http://dx.doi.org/10.1086/160960}{CrossRef}]

\bibitem[Rodr{\'\i}guez-Gil, et al.(2010)]{rod10}
Rodr{\'\i}guez-Gil, P.; Santander-Garc{\'\i}a, M.; Knigge, C.; Corradi, R.L.M.;
G{\"a}nsicke, B.T.; Barlow, M.J.; Drake,~J.J.; Drew, J.; Miszalski, B.; Napiwotzki, R.; et al. The orbital period of V458 Vulpeculae, a post-double
common-envelope nova. \textit{Mon. Not. R. Astron. Soc.} \textbf{2010},
\textit{407}, L21. [\href{http://dx.doi.org/10.1111/j.1745-3933.2010.00895.x}{CrossRef}]

\bibitem[Tovmassian, et al.(2010)]{tov10}
Tovmassian, G.; Yungelson, L.; Rauch, T.; Suleimanov, V.; Napiwotzki, R.; Stasi{\'n}ska, G.; Tomsick,~J.; Wilms,~J.; Morisset, C.; Pena, M.;
et al. The Double-degenerate Nucleus of the Planetary Nebula TS 01: A Close Binary Evolution
Showcase.  \textit{Astrophys. J.} \textbf{2010}, \textit{714}, 178. [\href{http://dx.doi.org/10.1088/0004-637X/714/1/178}{CrossRef}]

\bibitem[Bruch, et al.(2001)]{bru01}
Bruch, A.; Vaz, L.P.R.;  Diaz, M.P. An analysis of the light curve of the post common envelope
binary MT Serpentis. \textit{Astron. Astrophys.} \textbf{2001}, \textit{377}, 898--910,
doi:10.1051/0004-6361:20011092. [\href{http://dx.doi.org/10.1051/0004-6361:20011092}{CrossRef}]

\bibitem[Shimanskii, et al.(2008b)]{shi08b}
Shimanskii, V.V.; Borisov, N.V.; Sakhibullin, N.A.; Sheveleva, D.V. MT Ser, a binary blue
subdwarf. \textit{Astron.~Rep.} \textbf{2008}, \textit{52}, 479--486. [\href{http://dx.doi.org/10.1134/S106377290806005X}{CrossRef}]

\bibitem[Drilling(1985)]{dri85}
Drilling, J.S. LSS 2018: A double-lined spectroscopic binary central star with an extremely large
reflection effect.  \textit{Astrophys. J.} \textbf{1985}, \textit{294}, L107. [\href{http://dx.doi.org/10.1086/184519}{CrossRef}]

\bibitem[Hilditch, et al.(1996)]{hil96}
Hilditch, R.W.; Harries, T.J.; Hill, G. On the reflection effect in three sdOB binary stars.
\textit{Mon. Not. R. Astron.~Soc.} \textbf{1996}, \textit{279}, 1380--1392,
doi:10.1093/mnras/279.4.1380. [\href{http://dx.doi.org/10.1093/mnras/279.4.1380}{CrossRef}]

\bibitem[Af{\c{s}}ar and Ibanoglu(2008)]{afs08}
Af{\c{s}}ar, M.; Ibanoglu, C. Two-colour photometry of the binary planetary nebula nuclei
UU Sagitte and V477 Lyrae: Oversized secondaries in post-common-envelope
binaries. \textit{Mon. Not. R. Astron. Soc.} \textbf{2008}, \textit{391}, 802--814. [\href{http://dx.doi.org/10.1111/j.1365-2966.2008.13927.x}{CrossRef}]

\bibitem[Hillwig, et al.(2010)]{hil10}
Hillwig, T.C.; Bond, H.E.; Af{\c{s}}ar, M.; De Marco, O. Binary Central Stars of Planetary
Nebulae Discovered through Photometric Variability. II. Modeling the Central
Stars of NGC 6026 and NGC 6337. \textit{Astron. J.} \textbf{2010}, \textit{140}, 319,
doi:10.1088/0004-6256/140/2/319. [\href{http://dx.doi.org/10.1088/0004-6256/140/2/319}{CrossRef}]

\bibitem[Shimanskii, et al.(2004)]{shi04}
Shimanskii, V.V.; Borisov, N.V.; Sakhibullin, N.A.; Surkov, A.E. The Nature of the Unique
Precataclysmic Variable V664 Cas with Two-Peaked Balmer Lines in Its Spectrum.
\textit{Astron. Rep.} \textbf{2004}, \textit{48}, 563--576. [\href{http://dx.doi.org/10.1134/1.1777274}{CrossRef}]

\bibitem[Hillwig, et al.(2016)]{hil16a}
Hillwig, T.C.; Bond, H.E.; Frew, D.J.; Schaub, S.C.;  Bodman, E.H.L. Binary Central Stars of
Planetary Nebulae Discovered through Photometric Variability. IV. The Central
Stars of HaTr 4 and Hf 2-2. \textit{Astron. J.} \textbf{2016}, \textit{152}, 34,
doi:10.3847/0004-6256/152/2/34. [\href{http://dx.doi.org/10.3847/0004-6256/152/2/34}{CrossRef}]

\bibitem[Shimanskii, et al.(2008c)]{shi08c}
Shimanskii, V.V.; Borisov, N.V.; Pozdnyakova, S.A.; Bikmaev, I.F.; Vlasyuk, V.V.; Sakhibullin, N.A.; Spiridonova, O.I.
Fundamental parameters of BE UMa revised. \textit{Astron. Rep.} \textbf{2008}, \textit{52}, 558--575,
doi:10.1134/S1063772908070056. [\href{http://dx.doi.org/10.1134/S1063772908070056}{CrossRef}]

\bibitem[Ivanova, et al.(2013)]{iva13}
Ivanova, N.; Justham, S.; Chen, X.; De Marco, O.; Fryer, C.L.; Gaburov, E.; Ge, H.; Glebbeek, E.; Han, Z.; \mbox{Li, X.D.;
et al.} Common envelope evolution:
Where we stand and how we can move forward. \textit{\mbox{Astron. Astrophys. Rev.}}
\textbf{2013}, \textit{21}, 59. [\href{http://dx.doi.org/10.1007/s00159-013-0059-2}{CrossRef}]

\end{thebibliography}
\end{document}